\documentclass[12pt]{iopart}
\usepackage{iopams}
\usepackage{color}
\usepackage{url}

\expandafter\let\csname equation*\endcsname\relax
\expandafter\let\csname endequation*\endcsname\relax
\usepackage{amsmath}
\usepackage{graphicx}
\usepackage[usenames,dvipsnames]{xcolor}
\usepackage{multirow}
\usepackage{harvard}

\begin{document}

\title[Averaged cross sections for thermal mixtures of $\alpha$-Alanine conformers]{Averaged electron collision cross sections for thermal mixtures of $\alpha$-Alanine conformers in the  gas phase}

\author{Milton M. Fujimoto$^{1,3}$, Erik V. R. de Lima$^{1}$ and Jonathan Tennyson$^{2}$}

\address{$^1$Departamento de F\'{\i}sica, Universidade Federal do Paran\'a,
        81531-990 Curitiba, PR, Brazil}

\address{$^2$Department of Physics \& Astronomy, University College London,
Gower St., London, WC1E 6BT, UK}

\ead{$^3$milton@fisica.ufpr.br}

\begin{abstract}

A theoretical study of elastic electron collisions with 9
conformers of the gas-phase amino acid $\alpha$-alanine
(CH$_3$CH(NH$_2$)COOH) is performed. The eigenphase sums, resonance
features, differential and integral cross sections are computed for
each individual conformer. Resonance positions for the low-energy
$\pi^*$ shape resonance are found to vary from 2.6 eV to 3.1 eV
and the resonance widths from 0.3 eV to 0.5~eV. Averaged cross
sections for thermal mixtures of the 9 conformers are presented.
Both theoretical and experimental population
ratios are considered. Thermally-averaged cross sections
obtained using the best
theoretical estimates give reasonable agreement with the observed thermal cross sections. Excited conformers IIA and IIB make a large contribution to this
average due to their large permanent dipole moments.

\end{abstract}

\maketitle

\section{Introduction}

\begin{figure}[htb!]
 \centering
\includegraphics[width=8.3cm]{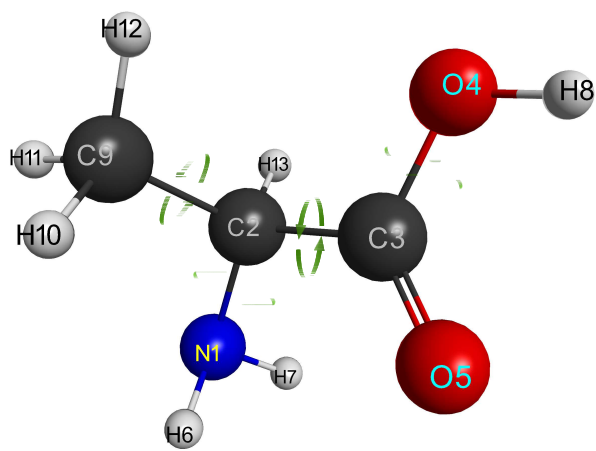}
\vspace{0.5cm}
 \caption{Structure of  {\it L}-$\alpha$-alanine. This amino acid is very 
flexible molecule, the rotation of various bonds generates many 
different conformers of $\alpha$-alanine amino acid.
(Figure generated with help of MacMolplt software~\cite{mcmolplt})}
  \label{fig.alanine}
\end{figure}

$\alpha$-Alanine, (CH$_3$CH(NH$_2$)COOH), is the second simplest
$\alpha$-amino acid found in the nature, and it is a building block of
proteins~\cite{Csaszar1999}.  $\alpha$-Alanine is a solid at room temperature 
and its crystal has a 
zwitterion form~\cite{Simpson,Dunitz}. However,
$\alpha$-alanine vaporizes at temperatures
near 120 $^\circ$C and it is well-known that in the gas phase it exists as a
mixture of charge-neutral conformers \cite{Godfrey,Blanco}. 
As is often
observed for amino acids~\cite{Iijima}, $\alpha$-alanine 
is a very
flexible molecule and it exists as a  great variety of 
conformers; Figure~\ref{fig.alanine} illustrates possible rotations
of $\alpha$-alanine. In the gas phase, intramolecular interactions are
important for stabilizing different spatial
arrangements of atoms. 

There is a lot of work on the alanine conformers in gas phase.
Initially, \citeasnoun{Iijima2} used electron diffraction to study a
gas phase sample. They assumed that the vapour of $\alpha$-alanine was
basically composed of a single conformer which had a high rotational
barrier around the C2-C3 bonds (following the carbon numbers in the
Figure~\ref{fig.alanine}), which therefore hindered conversion to
other conformers. In contrast, subsequently, many experimental and
theoretical studies showed that in the gas phase, several neutral
conformers of $\alpha$-alanine can coexist at temperatures for which
the experiments studies are performed, because their total energies
only differ by around 0.1 eV. \citeasnoun{Godfrey} analysed the
observed rotational spectra of alanine in the gas phase aided by
theoretical calculations performed at the self-consistent field
restricted Hartree-Fock (RHF) level of theory using a 6-31G* Gaussian
type orbital (GTO) basis set.  They concluded that they observed two
conformers, which were identified as I and III. The relative
concentration observed in their Stark-modulated free-expansion jet at
255 $^\circ$C was estimated to be 8:1 (I:III).  \citeasnoun{Cao}
optimized energies and geometries for 13 conformations at various
levels of theory, RHF/6-31G** up to second-order M{\o}ller-Plesset
perturbation theory \cite{Moller}: MP2/6-311G*, and partially
confirmed the previous results of Godfrey {\it et al}.  However, Cao
{\it et al} found that at the MP2 level conformer V was predicted to
have the same set of experimental rotational constants and dipole
moment of conformer I; they therefore concluded that was not possible
to unambiguously assign which conformer were being detected by
microwave spectroscopy.  \citeasnoun{Gronert} also searched for stable
structure of conformers of alanine as well as serine and cysteine.
They started the study at the semi-empirical Austin Model 1 (AM1)
\cite{Dewar} level and then refined the geometries using RHF/6-31G* up
to MP2/6-31+G* level of theory. For alanine, Gronert and O'Hair found
10 conformers; they reported rotational constants and dipole moments
for all of them.  Finally, \citeasnoun{Csaszar} performed {\it ab
  initio} calculations at different levels of theory including high
level of correlation such as the so-called 'gold standard' coupled
clusters single double (perturnative triples), CCSD(T).  As
$\alpha$-alanine is a very flexible molecule, intramolecular
interactions, due to hydrogen bonding and steric effects, are
important so the choice of appropriate diffuse and polarized basis
functions are important for predictions of the correct geometries and
relative energies. By analysis of minima on the potential energy
surface for neutral $\alpha$-alanine, Cs\'asz\'ar found 13 conformers
and reported accurate geometries, various properties and spectroscopic
constants of simulation of rotational and vibrational spectra.
Cs\'asz\'ar's theoretical relative energy predictions include
zero-point vibrational energy corrections, differ significantly from
the experimental limits given by \citeasnoun{Godfrey} in their
combined rotational study and structural study. This indicated the
need for an improved description of the gas-phase electron diffraction
of $\alpha$-alanine.  Cs\'asz\'ar's structural results support the
molecular constants measured for two low relative energy conformers. A
more recent, {\it ab initio}, focal point analysis by
\citeasnoun{Jaeger} essentially confirmed Cs\'asz\'ar's results.

Based on Cs\'asz\'ar relative energies and considering a temperature
of 500~K, one can expect the presence of several conformers in
appreciable relative concentration in the gas phase. 
\citeasnoun{Kaschner} used density functional theory (DFT) to
compute relative energies and geometries of some amino acids and
small oligopeptides. These authors reported results for 6 alanine
conformers and observed that the LDA (local-density approximation)
approach overestimates the strength of hydrogen bonds and cannot
reproduce the correct ordering compared with {\it ab initio}
calculations.  Kaschner and Hohl's
calculations confirmed that, in the gas phase, glycine and alanine prefer
neutral than zwitterionic form.  \citeasnoun{Lessari}
developed a method which combined laser ablation with rotational spectroscopy in a
supersonic jet; this allowed them to obtain structural
information for each individual conformers present
in the adiabatic expansion. This technique is most useful
for low volatility organic compounds as it avoids using of thermal
heating where the amino acid may decompose thermally  before
reaching their melting point.  \citeasnoun{Blanco} using
laser-ablation molecular-beam Fourier transform microwave spectroscopy
(LA-MB-FTMW) studied the rotational spectrum of neutral alanine in a
supersonic jet. They observed the two lowest energies conformers of
Alanine I and IIA. Their study failed to observed the presence of 
conformer IIB or any type III conformers in the supersonic jet using
using Ne or He as carrier gases.  Blanco {\it et al} suggested that 
collisional relaxation could convert the absent conformers in the
most stable IIA and I forms because of the low-energy barriers for
this process. They found 
the relative abundances between I and IIA conformers 
for post-expansion was $N_I$ /$N_{IIA}$ = 4, assuming that
both conformers are in the lowest vibrational state. This
finding is in good
agreement with their predicted equilibrium of the relative populations at 298 K
but contrasts with the population ratio obtained by \citeasnoun{Godfrey}($N_I$ /$N_{IIA}$ = 8).  
\citeasnoun{Balabin} performed a jet-cooled
Raman spectroscopic study and identified two of the missing conformers of
alanine, IIB and IIIA. His relative population for $N_I$ /$N_{IIA}$
= 5 which  agrees with Blanco {\it et al}. Balabin
suggested that the detection of these two conformers in
significant concentration was related to his different experimental set up
and also due to selective collisional relaxation processes associated
with low interconversion barriers in the jet. \citeasnoun{Feyer}
recorded core level X-ray photoemission spectra (XPS) and near edge X-ray
absorption fine structure (NEXAFS) spectra of alanine in the gas phase
measured at the carbon, nitrogen, and oxygen K edges.
With the help of theoretical calculations they determined the population
ratio for conformers I and II of alanine, 0.78:0.21, respectively,
which are also in good agreement with previous results of Blanco {\it et
  al} and Balabin. \citeasnoun{Powis} studied photoelectron spectra of gas-phase alanine using synchrotron radiation. They
concluded that their results are consistent with the suggestion that
conformer I is the predominant structure; when they tried to include
conformers II and III in any proportion in their analysis they did not
improve agreement with their experimental results. Although the I-III
structures that they optimized by DFT B3LYP/6-31G** calculations had very
close relative energies so that they would be expected in equal thermal
population. \citeasnoun{Farrokhpour} in a 
combined theoretical
and experimental study estimated the population ratio of the four
lowest energy conformers of alanine from the experimental
photoelectron spectrum supported by high level theoretical
calculations. After fitting the spectra the relative population that
they obtained for the conformers were I = 70\%, IIA = 14\%, IIB = 10\%
and III = 6\% at 403 K.

 \citeasnoun{Bazso} measured near (NIR) and 
mid-infrared (MIR) spectra of $\alpha$-alanine isolated in matrices of Ar, 
Kr and N$_2$ at low temperature (8-14 K).  Upon short 
irradiation with NIR laser light at the first O-H stretching overtone 
band of conformer I, a short-lived conformer VI was observed, which 
decayed by H-atom tunnelling at 12 K. The half-lives were measured for 
each matrix. They also observed that when conformer I is exposed to 
prolonged irradiation it is transformed into conformer IIa, but conformer 
IIb was not observed. Their observations are  consistent with an 
extremely low barrier to interconversion of 
IIb $\rightarrow$ IIa and it was 
not expected that higher energy conformers could be isolated in a matrix 
at around 10 K. In contrast with glycine, they did not observed conformer IIIb because 
of the very low IIIb $\rightarrow$ I barrier height in alanine. 
Note that the 
nomenclature of conformers "a" and "b" used by \citeasnoun{Bazso} 
differs from the "A" and "B" conformer designations of \citeasnoun{Csaszar}).

The discovery that DNA molecules could be severely damaged by secondary
electrons with energies less than 20 eV, has led to increased 
research into  electron scattering by biological
molecules. The DNA damage is caused when the low-energy electron is
captured by an unoccupied orbital of a DNA constituent molecule creating a
transient anion or resonant state which can then decay into
negative and neutral fragments. Depending on what the type of DNA
constituent (such as, amino acids, sugar, nucleobase, or phosphate
moieties, etc) absorbs the electron, this process can  lead to single or double
strand brakes in the DNA molecule \cite{Boudaiffa}.

Conversely, there are relatively few works on electron collisions
cross sections with alanine in the gas phase. 
\citeasnoun{Marinkovic}, to our knowledge, were the first to measure
elastic cross sections for alanine in the range from 40 eV to 80 eV.
In a combined theoretical and experimental study, they compared their
measured data with calculated results from a corrected form of the
independent-atom method (IAM), known as the SCAR (Screen Corrected
Additivity Rule) procedure and obtained good agreement between them.
They present theoretical results for electron impact energies ranging 
from 1 to 10,000 eV.
\citeasnoun{Panosetti} focused on modelling resonance
fragmentation; they made a systematic study involving glycine,
alanine, proline and valine. Panosetti {\it et al} reported the
elastic integral cross sections (ICS) for these molecules at the
equilibrium geometry and observed resonance near 3 eV for all the
amino acids studied. These resonances are important for the
dissociative electron attachment (DEA) process which is the main
low-energy electron-collision destruction mechanism for neutral
molecules \cite{Aflatooni,Ptasinska}. \citeasnoun{Fujimoto} reported elastic
differential cross sections (DCS) and ICS for two conformers of alanine and 
observed differences in the
resonance positions between them.

In this paper, we report a theoretical study comparing elastic cross
sections for the 9 lowest-energy conformers of $\alpha$-alanine in the
gas phase for the energies ranging from 1 to 10 eV. We present
averaged-cross sections which take into account relative populations
of different conformers.  We believe that these should be more
reliable results for comparison with measured data at temperatures
where the gas phase molecule is experimentally accessible.

The organization of article is as follows: Section 2 presents an
outline of the theory and some details of the calculations are
provided in Section 3. Section 4 presents and discusses our calculated
data; this is followed by a summary of conclusions.

\section{Calculations}

\subsection{Geometry of Conformers of $\alpha$-alanine }
\label{sec:2}

In this work, we  use the nomenclature and geometry of conformers 
given in Table 2 of \citeasnoun{Csaszar}. 
Cs\'asz\'ar describes 
13 geometrical structures of $\alpha$-alanine corresponding to minima on the 
potential energy surface. We have used Cs\'asz\'ar's geometric parameters 
which were optimized at MP2/6-311++G** level; these parameters are given
in the supplementary material.

We aim to compute cross sections that can be compared with
experimental data, so we need to consider that in crossed-beam
experiments $\alpha$-alanine is sublimed by heating to 400 -- 500 K,
as it is solid at room temperature. Initially, we started to work with
the 13 conformers of alanine given by Cs\'asz\'ar.  However, below we
limit our study to only the 9 lowest energy conformers, I to VB (see
Table 1 of \citeasnoun{Csaszar}), because their Boltzmann population
ratios indicate that they dominate the population at 
temperatures at which the experiment are conducted.

\subsection{The R-matrix method}
\label{sec:22}

The R-matrix method used in this paper is the UKRMol implementation
\cite{jt518} of the UK molecular R-matrix 
codes which is describe in details elsewhere by \citeasnoun{Gillan} and 
\citeasnoun{Tennyson2}. 
\citeasnoun{Fujimoto} performed
 R-matrix calculations on two 
alanine conformers and present a detailed study 
of how this was done. Here we  
use the same approach to calculate cross sections for 9 conformers; 
the reader is  referred to the previous calculation for further details.

Here we just present an brief summary of the R-matrix method 
and calculations. In 
this method, space is split in two regions: the inner and outer region. The 
inner region is a sphere of radius $a$ around the target molecule
centre-of-mass.  The electronic density of 
the target molecule is considered, in practice, to be completely inside
this region. Here a 
radius of $a=10 a_0$ was used. The wave function of (N+1)-electron
system inside 
the sphere is given by
\begin{eqnarray}
\Psi_k^{N+1} (x_1 \ldots x_{N+1} )&=& {\cal A} \sum_{ij} a_{ijk} \phi_i^N (x_1
\ldots x_N ) u_{ij}
(x_{N+1} )   \nonumber \\
&+& \sum_i b_{ik} \chi_i^{N+1} (x_1 \ldots x_{N+1})
\label{eq1}
\end{eqnarray}
where $\phi_i^N$ are the wave functions of the target in the $i^{\rm th}$ state
and $u_{ij}$  is the wave function of the continuum electron which is expanded 
in a partial wave expansion
up to some maximum value of  $\ell$, $\ell_{\rm max}$;  $\cal A$ is an 
antisymmetrization operator as the (N+1)-electrons are indistinguishable in the inner-region electrons. The second summation in eq. (\ref{eq1}) contains 
configurations $\chi_i^{N+1}$ which are included to relax the constraint of
orthogonalization between scattering and target orbitals of the same symmetry,
and to allow for target polarization effects.
$a_{ijk}$ and $b_{ik}$ are the variationally-optimized coefficients of 
expansions.

The polarization effects, which are related to the distortion of target 
electronic density, are calculated when $\chi_i^{N+1}$ configurations are 
included to allow the relaxation of target bonded orbitals. This level of
calculation is called SEP (static-exchange-polarization). 
SEP calculations are known to provide a good practical
method of converging polarisation effects in low-energy electron
molecule collisions and, in particular, of providing
reliable predictions of resonance parameters in this region \cite{vrs08,jt533}.
Following tests by \citeasnoun{Fujimoto}, 30 virtual orbitals
are used to generate  $\chi_i^{N+1}$ configurations. 

The outer region, we solve a coupled second-order differential
equation for the continuum electron, it is a one-particle problem. The
R-matrix matches the inner and outer solutions, propagate to large $r$
and then uses the K-matrix to calculate scattering observables.  The
partial wave expansion includes partial waves explicitly up to
$\ell_{\rm max}$ ($\ell_{\rm max}$=4) and is then completed using a Born
closure procedure \cite{Morrison,Gianturco,Padial}. This takes into
account the long-range dipole interactions as all of alanine
conformers have permanent dipole moment. The code POLYDCS \cite{Sanna}
is employed to calculate rotational transitions and the rotational constants
used for each conformers are reported in supplementary material.
The rotationally-unresolved elastic differential cross sections is 
obtained by the sum over rotational transitions until the convergence.
Here we limit our partial wave expansion to  $\ell_{\rm max} \leq 5$;
our previous study on  alanine
\cite{Fujimoto} tested adding the 
$\ell_{\rm max}$=4 partial wave. However, once the Born closure (which allows
for the contribution of many partial waves particularly at low angles) is
performed, 
the results were essentially unchanged except that the
$\ell_{\rm max} \leq 5$ calculations showed 
unphysical behaviour in the DCS for angles higher than 150$^\circ$ when Born 
closure procedure was used.

\section{Results and discussion}
\label{sec:3}

In this section, we present our  results including eigenphase sums, 
resonance positions, DCS and ICS for elastic electron collisions with 
alanine molecules in the range of 1 to 10 eV. All the results were 
calculated in SEP level with 30 virtual orbitals.

\subsection{Target properties and population}

All our target wavefunctions for the 9 conformers are described at the
RHF/6-311+G* level and the relative energies compared with conformer I
are around only 0.1 eV. Table~\ref{tab:2} reports key data on target
conformers which are important for the calculations below. Target
wavefunctions and molecular properties were obtained using the MOLPRO
Quantum Chemistry package \cite{Werner}.

Table~\ref{tab:2} gives our relative energy and dipole moment calculated
at the RHF/6-311+G* level for all 9 conformers using geometries from
\citeasnoun{Csaszar} optimized at MP2/6-311++G** level.  
The Boltzmann relative populations are
presented for two levels of theory: RHF (where our RHF relative energies are
used), CCSD(T) (where relative energies calculated  with CCSD(T) method by
\citeasnoun{Csaszar} are used) and a set 
of experimental
relative population deduced by \citeasnoun{Farrokhpour}.  

Experimental studies of processes including DEA,
photoelectron spectra and elastic electron scattering require
alanine to  vaporized. In these gas phase studies the temperature, in
general, ranges from about 400~K to 500~K. We observe that for a
given level of theory, RHF or CCSD(T), the relative population is not
significantly affected over this range of temperatures, however the CCSD(T)
population ratio is very different for conformers IIA and
IIB, compared with that calculated at the RHF level. A detailed
discussion of the difference in the total energies of $\alpha$-alanine
conformers at various levels of theory is given by
\citeasnoun{Csaszar}, who commented that RHF calculations
overestimate the relative energies compared to conformer I and
fails to predict them reliably. According to Cs\'asz\'ar, when the electron
correlation is included it stabilizes all conformers relative to 
conformer I, but not uniformly. 
As can be seen from Table~\ref{tab:2}, the CCSD(T) relative
population for conformer IIA, IIB and IIIB are very different compared
to those predicted at the RHF level. The set of experimental
relative population we considered were deduced by 
\citeasnoun{Farrokhpour} from the measured photoelectron spectra.
Their analysis found that it was only necessary to consider 4 conformers (I,
IIA, IIB and IIIA) in the molecular beam to fit their
observations. Comparing the CCSD(T) population ratio  
with the experimental set, we find good agreement for IIA, IIB and
IIIA, but  a larger discrepancy for conformer I. 
However we note that our analysis suggests that about 30~\%\ of the 
population should be in conformers not considered by Farrokhpour
{\it et al} and, given, that all populations are normalised, this
leads to their population of conformer I being larger than our
predicted one. The discrepancies and the absence of some conformers may well be
related to the conversion from one conformer to another 
due to the very low barrier heights involved \cite{Blanco} or decay by H-atom 
tunnelling~\cite{Bazso}. 

\begin{table}
\caption{Relative energy(RE), dipole moment and relative
population in \% of molecules for 9 $\alpha$-alanine conformers at two
temperatures, 403 K and 500 K.}
\label{tab:2}       
\smallskip
\begin{tabular}{crcrrrrr}
\hline
 \multicolumn{3}{ c}{ } &\multicolumn{5}{c}{Relative Population, \% } \\ 
\cline{4-8}
 \multicolumn{3}{ c}{ }  &\multicolumn{2}{c}{RHF$^{\rm a}$}&\multicolumn{2}{c}{CCSD(T)$^{\rm b}$}& Expt$^{\rm c}$ \\ 
\hline\noalign{\smallskip}
\multicolumn{1}{c}{Conformer} & RE$^{\rm a}$ ($cm^{-1}$) & Dipole (D) & 403 K & 500 K & 403 K & 500 K & 403 K \\
\hline\noalign{\smallskip}
\multicolumn{1}{c}{I}   & 0    & 1.35 & 50.2 & 42.0 & 35.7 & 30.0 & 70.0 \\
\multicolumn{1}{c}{IIA} & 1126 & 6.10 & 0.9 & 1.6 & 16.1 & 15.8 & 14.0 \\
\multicolumn{1}{c}{IIB} & 1156 & 6.15 & 0.8 & 1.5 & 16.9 & 16.4 & 10.0 \\
\multicolumn{1}{c}{IIIA} & 584 & 1.77 & 6.2 & 7.8 & 6.7 & 7.8 & 6.0 \\
\multicolumn{1}{c}{IIIB} & 214 & 1.55 & 23.3 & 22.6 & 7.5 & 8.5 &  \\
\multicolumn{1}{c}{IVA} & 594 & 2.36 & 6.0 & 7.6 & 6.1 & 7.3 &  \\
\multicolumn{1}{c}{IVB} & 633 & 2.32 & 5.2 & 6.8 & 5.1 & 6.2 &  \\
\multicolumn{1}{c}{VA}  & 666 & 2.66 & 4.7 & 6.2 & 3.9 & 5.1 &  \\
\multicolumn{1}{c}{VB}  & 832 & 2.88 & 2.6 & 3.8 & 2.0 & 2.9 &  \\
\noalign{\smallskip}\hline
\end{tabular}
\begin{itemize}
\item[] $^{\rm a}$ Our relative energies calculated at the RHF level.
\item[] $^{\rm b}$ Relative energies from CCSD(T) calculations by 
\citeasnoun{Csaszar}.
\item[] $^{\rm c}$  Experimental data from \citeasnoun{Farrokhpour}.
\end{itemize}
\end{table}

The relative energies and dipole moments presented in Table 1 are 
calculated in RHF level  are slightly different from those reported by 
Cs\'asz\'ar for his RHF/6-311++G** level
calculations \cite{Csaszar} because we used 
geometries optimized by Cs\'asz\'ar in MP2 level and a 6-311+G* 
basis set . Despite this small differences we can expect similar results. 
Comparing our total energies against the results 
(not shown) of \citeasnoun{Csaszar} for each conformer, the differences are 
around 0.03\%. 
The maximum relative difference between our calculated dipole moments and those 
of Cs\'asz\'ar is 10.8\% for conformer IIIB and about 9\% for 
conformers IIA and IIB. The difference in dipole moment for conformer I 
is 3.8\% (1.4D) compared with Cs\'asz\'ar and 25,2\% (1.8D) compared with 
the experimental 
results of \citeasnoun{Godfrey}. Conformer IIB has a dipole moment 
18\% higher than experimental results (5.13 D) \cite{Godfrey}.

\subsection{Eigenphase sums and Resonances}
\label{sec:3.1}

\begin{figure}[htb!]
 \centering
\includegraphics[width=11cm]{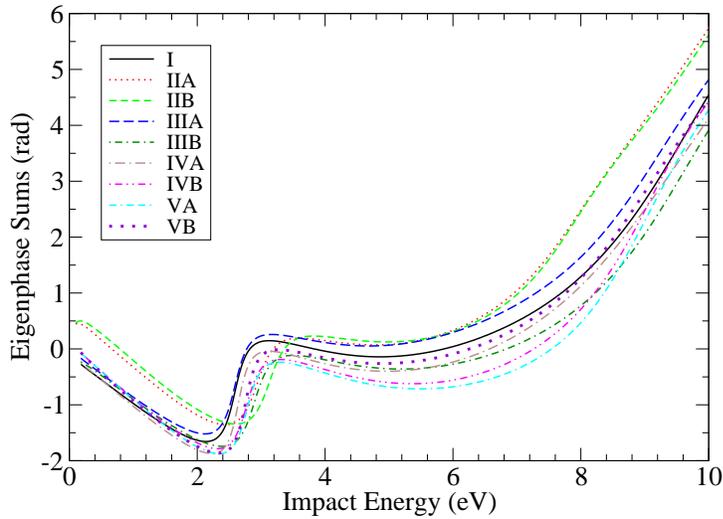}
 \caption{Eigenphase sums for 9 conformers of $\alpha$-alanine 
computed in the SEP level.}
  \label{fig.2-eigen}
\end{figure}

Figure~\ref{fig.2-eigen} shows the eigenphase sums for 9 conformers of
$\alpha$-alanine. In all cases there is a sharp structure around 2.8
eV that is related to the electron capture at $\pi^*$ unoccupied
orbital of the carboxyl group as discussed
previously \cite{Aflatooni,Ptasinska}. A more detailed discussion of
this resonances is given by \citeasnoun{Fujimoto} and
the references cited therein.  Table~\ref{tab:1} presents our
resonance positions and widths which are fitted by Breit-Wigner
formula made by an automated fitting procedure \cite{Tennyson}. As
found by Fujimoto {\it et al}, our fits give a sharp lower-energy and
a broad higher-energy resonance.  Table~\ref{tab:1} shows that there
is a spread in the resonance positions and widths. The positions of
lower-energy resonance vary from near 2.6 eV up to 3.1 eV and the
widths vary from 0.3 eV to 0.5 eV. We would expect the positions of
these lower-energy resonance to be well described by our SEP model;
the resonance is formed by electron captured by the unoccupied $\pi^*$
orbitals of carboxyl group which is relatively straightforward to
model. The formation mechanism for the higher-energy resonance is not
well-established: it has been suggested that is a core-excited
resonance \cite{Tashiro} or a shape resonance 
associated with the $\sigma^*$ unoccupied orbital of OH group
\cite{Scheer}. These resonances are broader: the positions of higher energy
resonance ranges from 8.0 eV to 9.8 eV and the widths from 1.8 eV
to 3.6 eV.

\begin{table}
\center
\noindent
\caption{Resonance parameters for 9 $\alpha$-alanine conformers: positions (widths) in eV.}
\label{tab:1}       
\begin{tabular}{cccc}
\noalign{\smallskip}
\hline\noalign{\smallskip}
Conformer & \multicolumn{3}{c}{Resonances: position (width)}  \\
\noalign{\smallskip}\hline\noalign{\smallskip}
I   & 2.59(0.35) &      & 9.70(2.19) \\
IIA & 2.95(0.50) & 8.16(2.18) & 9.60(2.65) \\
IIB & 3.14(0.52) & 8.03(1.80) & 9.77(2.41) \\
IIIA & 2.59(0.38) &      & 9.32(2.32) \\
IIIB & 2.90(0.46) &      & 9.67(2.16) \\
IVA  & 2.64(0.31) &      & 9.74(3.60) \\
IVB  & 2.78(0.42) &      & 9.10(1.58) \\
VA   & 2.78(0.39) &      & 9.06(1.76) \\
VB   & 2.76(0.32) &      & 9.12(2.39) \\
\noalign{\smallskip}\hline
\end{tabular}
\end{table}

\subsection{Cross sections}
\label{sec:3.2}

\begin{figure}[htb!]
 \centering
\includegraphics[width=14cm]{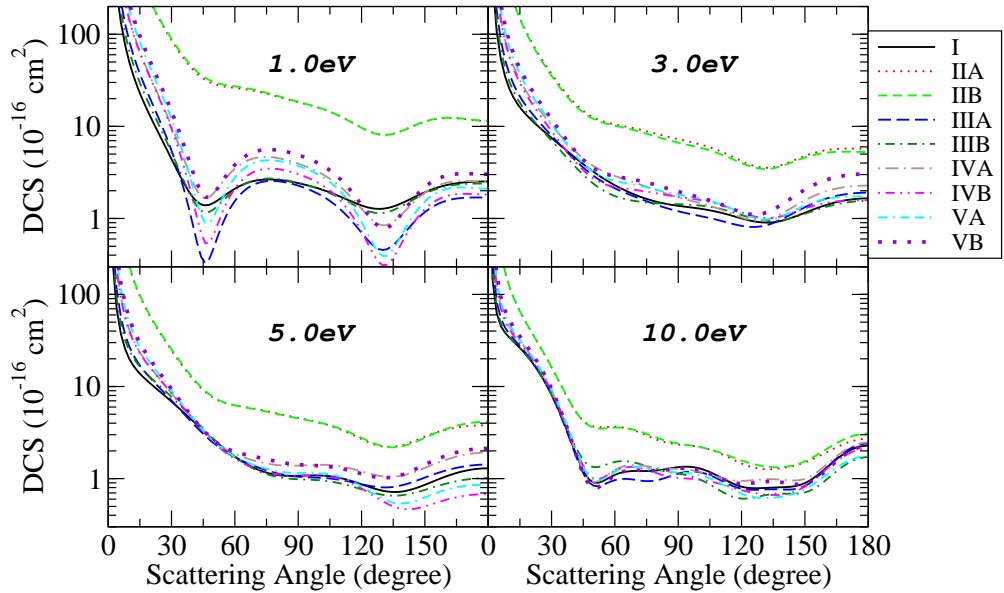}
 \caption{Elastic DCS for electron collision with $\alpha$-
alanine conformers for impact energies of 1, 3, 5 and 10 eV. Results for  
SEP calculations including a Born correction.}
  \label{fig.3-dcs}
\end{figure}

Figure~\ref{fig.3-dcs} presents our DCS for 9 conformers of 
$\alpha$-alanine including 
polarization and completed with Born correction for 
collision energies 1, 3, 5, 
and 10 eV. Results calculated with steps of 1 eV from 1 to 10 eV
are given in the supplementary material. 
The DCS of all of conformers show similar behaviour and magnitudes 
between them with the exception of 
two conformers: IIA and IIB.  
Table~\ref{tab:2} shows that the main 
reason for the huge increase in the DCS of conformers IIA and IIB 
is due to their large dipole moments: 6.10 D and 6.15 D, 
respectively. The calculations are supplemented 
with Born
corrections to take into account higher partial waves beyond of
$\ell_{\rm max}$=4 which  
corresponds with including the long-range dipole interactions. 
This corrections depends on the square of the permanent dipole moment.
This means that 
in the case of 
conformers IIA and IIB the long-range interactions
contribute at least an order of magnitude more to the cross sections
than for the other conformers studied here.

\begin{figure}[htb!]
 \centering
\includegraphics[width=14cm]{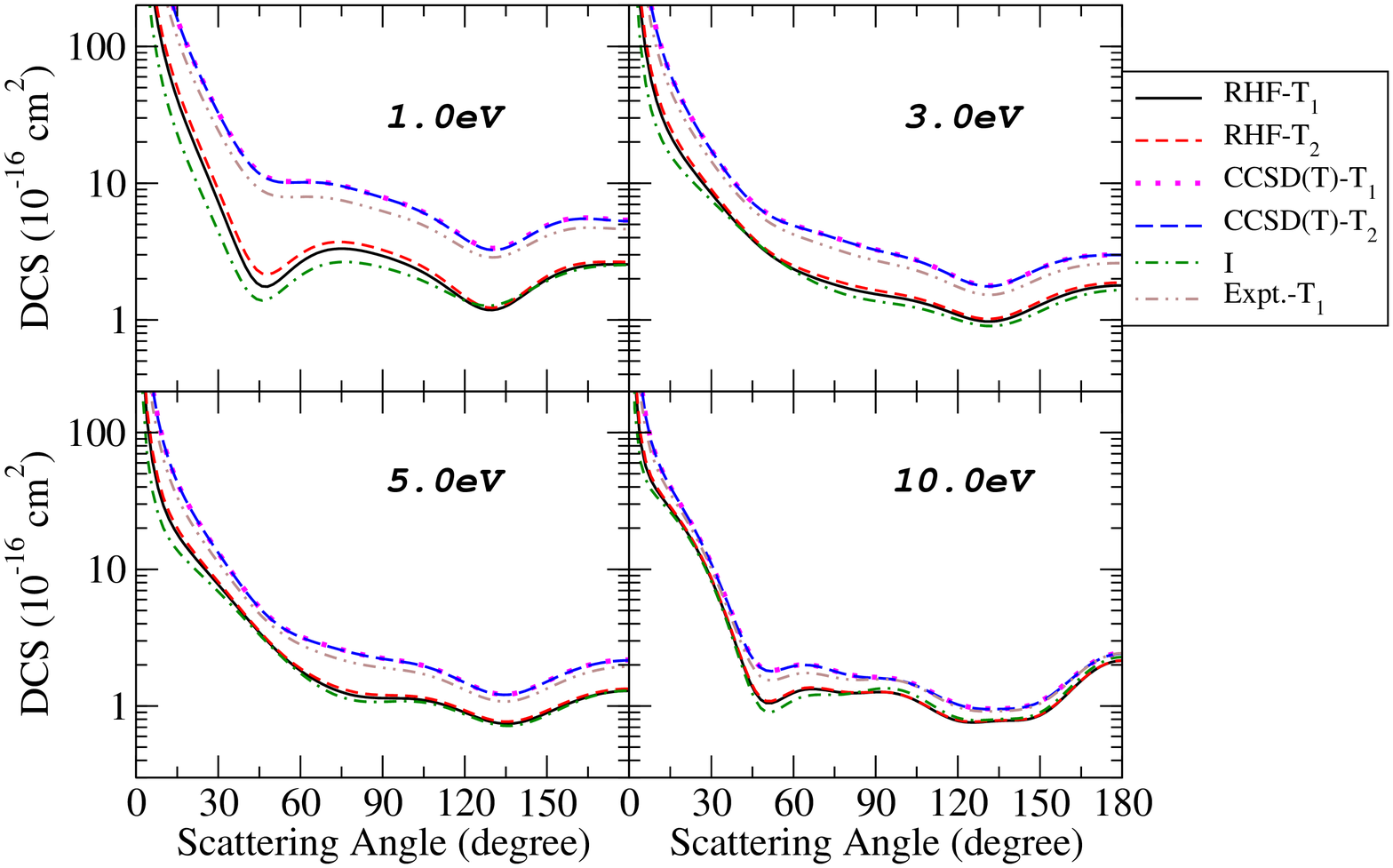}
 \caption{Comparison between averaged Born-corrected DCS using different sets of population ratio calculated with energies in RHF level, 
CCSD(T)  level for temperature of T$_{1}=403$ K and T$_2=500$ K and a 
set estimated from experimental data \cite{Farrokhpour} for T$_{1}=403$ K.
The results are presented for impact energies of 1, 3, 5 and 10 eV; a comparison
with conformer I is also presented.}
  \label{fig.4-dcsavg}
\end{figure}

Figure~\ref{fig.4-dcsavg} compares averaged DCSs calculated 
with the three different sets 
of Boltzmann population ratio given in Table~\ref{tab:2}. The averaged 
DCS were calculated by the sum of individual DCS weighted by 
population ratio, given by

\begin{eqnarray}
({\rm DCS})^{\rm avg}(T) &=&  \sum_{i} c_{i}(T) ({\rm DCS})_i 
\label{eq2}
\end{eqnarray}
where the $c_{i}(T)$ are the temperature-dependent population ratios 
and ${\rm(DCS)}_i$ is the SEP  
DCS of  conformer $i$. 
	
Table~\ref{tab:2} shows that for temperatures in the range 403~K to 500~K,
the population ratio does not change significantly. The consequence of
this can be seen in Figure~\ref{fig.4-dcsavg} which compares the
averaged-DCS calculated with population ratios computed using the same
model but for different temperatures. The differences are not so
distinct. Considering the RHF  
population ratios, the averaged cross section in
Figure~\ref{fig.4-dcsavg} for temperature of 403 K and 500 K are
fairly similar. At the RHF level  the averaged DCS are closer
to the DCS of lowest-energy conformer, conformer I, than the
averaged-DCS calculated using the CCSD(T) population ratio. In the
CCSD(T) case, the temperature again almost does not affect the
average because the cross sections of conformers IIA and IIB,
whose populations do not change significantly in the
temperature range considered, are dominant.

The averaged DCS calculated with CCSD(T) population ratios 
are larger at all angles than the RHF ones. This is because
the  CCSD(T) ratios lead to much larger populations of
  conformers IIA and IIB which have much bigger cross sections
due to their large permanent dipoles:
the combined CCSD(T)
population of these conformers over 30\% compared with around 2.5\%
for the RHF ratios. This shows that 
the thermally-averaged DCS, and by analogy other cross
sections, can be very
different from the DCS of the lowest-energy conformer I. 
Figure~\ref{fig.4-dcsavg} also  shows the  averaged
DCS based on the experimental population ratio reported by
Farrokhpour {\it et al}.  Comparing all results we
infer that if we want to compared theoretical cross sections of very
flexible molecules, such as, some amino acids, with correspondent
measured cross sections, we should calculate cross sections for those
conformers that are populated at temperatures at which
the experiment is conducted, and then evaluate the averaged cross
section.

\begin{figure}[htb!]
 \centering
\includegraphics[width=12cm]{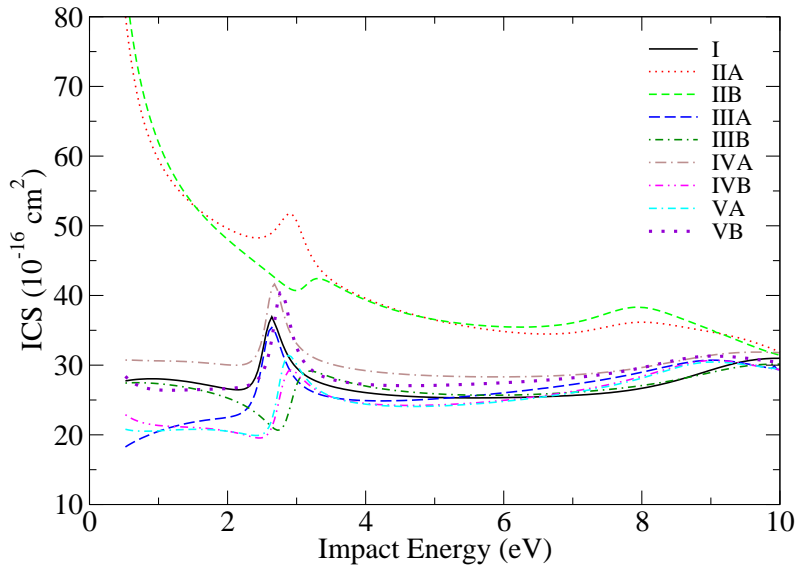}
 \caption{ICS for elastic electron scattering by 
9 $\alpha$-alanine conformers in SEP model not corrected with Born closure. 
For impact energies from 1 eV to 10 eV. The resonance features are shown.}
  \label{fig.5-icswb}
\end{figure}

Figure~\ref{fig.5-icswb} shows the ICS without including the Born
closure correction as we want to highlight the resonance features.
The lower-energy resonance peak can be seen for all conformers at
around 2.8 eV. A another broad resonance peak around 8 and 9 eV can
also be seen. The ICS for conformers IIA and IIB are greater in 
magnitude by around 40\% above  impact energies of 3.5 eV 
compared to all of the others
conformers; this is again due to the large dipole moments
of these conformers. If the long-range interactions are suppressed,
because the Born closure procedure is not considered, the ICS
at energies around 10 eV are very similar for all conformers,
indicating that at this electron scattering energy  the differences
between the geometries of the conformers are not so important in terms of
the short-range contribution to the
ICS.

\begin{figure}[htb!]
 \centering
\includegraphics[width=12cm]{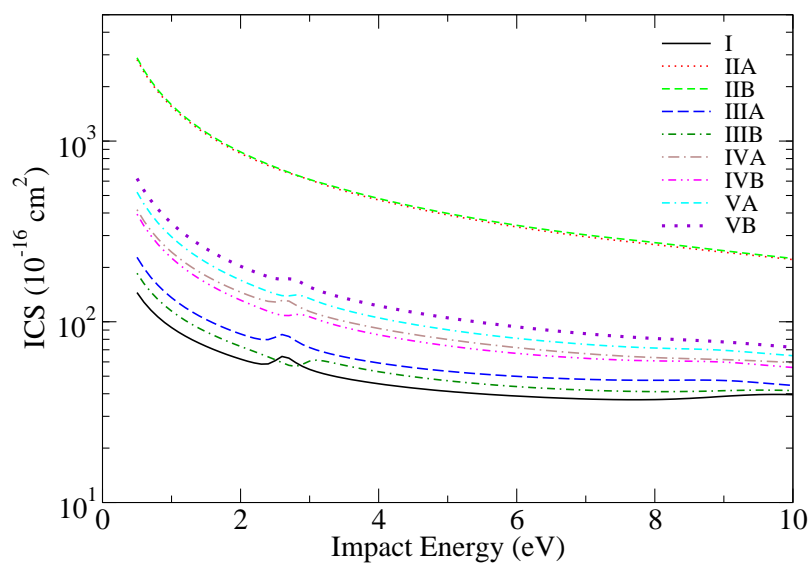}
 \caption{ICS for elastic electron scattering by 9 $\alpha$-alanine 
conformers calculated using a SEP model including Born correction.}
  \label{fig.6-icsbc}
\end{figure}

Figure~\ref{fig.6-icsbc} shows the ICS computed at the SEP level
including a Born closure correction. The cross sections
fall into three separate groups: one consisting of conformers I, IIIA
and IIIB whose dipole moments vary from 1.35 to 1.77 D. Another group
is composed of conformers VI-A,B and V-A,B whose dipoles ranges from
2.36 D to 2.88 D. The third group is formed by conformers IIA and IIB
which dipoles are around 6 D. Furthermore, these long-range interactions
effects dominate polarization effects for all conformers, and for
conformers IIA and IIB the resonance peak in the ICS was completely
washed out.

\begin{figure}[htb!]
 \centering
\includegraphics[width=12cm]{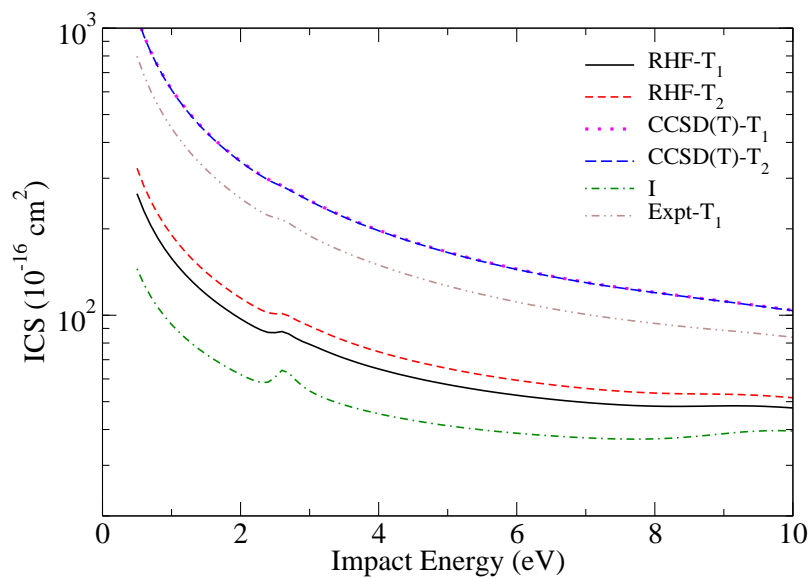}
 \caption{Averaged elastic ICS for electron scattering 
by $\alpha$-alanine conformers. Comparison of between different sets of
population ratio: RHF, CCSDT(CC) and experimental \cite{Farrokhpour} at  
temperatures of T$_{1}=403$ K and T$_2=500$ K. A comparison with conformer
I is also presented.}
  \label{fig.7-icsavg}
\end{figure}

Figure~\ref{fig.7-icsavg} reports the averaged ICS computed using SEP
and the Born closure correction for different population ratios (RHF,
CCSD(T) and Experimental), and at temperatures 403 K and 500~K. These
are compared with ICS for conformer I.  The averaged ICS calculated
with the RHF populations at 500~K shows a bigger ICS compared to 403
K, as expected as the relative contributions of conformers with
energies above  conformer I increases for higher temperatures, and this
particularly the cross sections of conformers IIA and IIB contribute
increases the average ICS. Even though the difference
between the two temperatures is
not large, the increase is around 14\%. The ICS calculated with RHF
populations is fairly similar to the ICS of conformer I
because the contributions due to conformers IIA and IIB are relatively
small. The averaged ICS calculated with the CCSD(T) population ratio are
bigger than that one calculated at the RHF level, as the relative
contributions of IIA and IIB conformers increases.  
The averaged ICS weighted with the experimental
population ratios are closer to  our averaged ICS
calculated with the CCSD(T) populations.  We note 
Cs\'asz\'ar affirmed that RHF fails to predict the
relative energies of $\alpha$-alanine correctly. Therefore
the ICS given by the CCSD(T) or experimental
populations represent our best predictions of the thermally averaged ICS.

\section{Conclusions}

In this work we report a theoretical study of elastic cross 
sections for electron collisions with nine gas-phase alanine conformers,
computed at the SEP level, in the range of 1 to 10 eV. 
The results show that the low-lying shape resonance parameters are relatively
unaffected by changes in conformer; similar results have been obtained
before for other systems with multiple minima \cite{jt399,jt487}.
However this in not true for the differential or integral cross sections 
where differences in dipole moments lead to large differences in
the computed cross sections between conformers. This suggests
that measured cross sections should depend on the relative population
of the conformers and temperature. 

Our results shows that for temperature ranging from 400~K to 500~K do
not affect significantly the magnitude of cross sections.  For this
range of temperature, considering relative energies from the
lowest-energy conformer I, the Boltzmann statistics suggest that 9
conformers should be significantly populated in the gas phase.  With
flexible molecules, such as some amino acids, computed cross sections
should be averaged over all conformers weighted by the expected
population ratio before comparing with the measured data. Our results
show that the averaged cross sections can be very different from those
calculated for just the most stable conformer I. This means that
averaged cross sections should give a more reliable to comparison with
measurements. In particular, if some conformer has a particularly
large cross section because, for example, it has a large dipole
moment, its contribution to the total cross sections will be very
dependent on its relative population, which will vary with
temperature.  Furthermore, the population ratios calculated using two
levels of theory, RHF and CCSD(T), to give the conformer energies are
significantly different.  The RHF calculations fail to predict the
relative energies correctly, especially for those molecules which have
intramolecular hydrogen bond, as is the case for amino acids. The
population ratio calculated using higher levels of theory which
include electron correlation, such as CCSD(T), give different
population ratios and consequently a different averaged cross section.
In this case the predicted population ratios for conformer IIA, IIB
and IIIA appear to be in good agreement with those deduced from
experimental. 

In conclusion, as the cross sections depend on the geometry of the
conformer or conformers considered, we recommend that for the authors
who will publish theoretical cross sections for large and flexible
molecules, such as amino acids, both specify which geometry was used for
their calculation, otherwise their results will not
be reproducible, and consider the effect of thermal
averaging when comparing with experiment results.

\section*{Acknowledgements}
We thank Attila Cs\'asz\'ar for helpful comments on our original manuscript.
M.M.F. acknowledges partial support from the Brazilian agency Conselho 
Nacional de Desenvolvimento Cient\'ifico e Tecnol\'ogico (CNPq) and 
E.V.R.L for a grant for UFPR-TN.

\section*{References}
 \bibliographystyle{jphysicsB}

\end{document}